\documentclass[%
 reprint,
superscriptaddress,
 amsmath,amssymb,
 aps,
]{revtex4-2}
\usepackage{graphicx}
\usepackage{amsmath}
\usepackage{amssymb}
\usepackage{color}
\usepackage{dcolumn}
\usepackage{bm}

\begin{document}
\title{Dynamical scaling behavior of the two-dimensional random singlet state in the random $Q$ model}

\author{Chen Peng}
\email{pengchen@ucas.ac.cn}
\affiliation{Kavli Institute for Theoretical Sciences and CAS Center for Excellence in Topological Quantum Computation, University of Chinese Academy of Sciences, Beijing 100190, China}
\author{Long Zhang}
\email{longzhang@ucas.ac.cn}
\affiliation{Kavli Institute for Theoretical Sciences and CAS Center for Excellence in Topological Quantum Computation, University of Chinese Academy of Sciences, Beijing 100190, China}

\date{\today}

\begin{abstract}
    In this work, we study the scaling relation of energy and length scales in the 2D random-singlet (RS) state of the random $Q$ model. To investigate the intrinsic energy scale of the spinon subsystem arising from the model, we develop a constrained subspace update algorithm within the framework of the stochastic series expansion method (SSE) to extract the singlet-triplet gap of the system. The 2D RS state exhibits scaling behavior similar to the infinite randomness fixed point (IRFP), at least within the length scales that we simulate. Furthermore, by rescaling the system size according to the strength of randomness, we observe that the data for the excitation gap and the width of the gap distribution collapse onto a single curve. This implies that the model with different strengths of randomness may correspond to the same fixed point.
\end{abstract}

\maketitle
\section{Introduction} 
In the early study of the interplay between quenched disorder and the quantum spin system, the so-called random singlet (RS) state was discovered in the 1D Heisenberg chain with random magnetic exchange strengths. The most intriguing phenomenon in the RS state is the power-law divergence of the heat capacity and the paramagnetic susceptibitliy as $T\rightarrow0$, which does not depend on the specific form of the initial randomness distribution of the Heisenberg chain. This suggests a universal effective power-law distribution of antiferromagnetic (AF) exchange energies at low temperatures. This 1D RS phase can be treated using the strong disorder renormalization group (SDRG) and Fisher found this paramagnetic RS phase corresponds to the exact infinite randomness fixed point\cite[]{Fisher.PRB.1994} (IRFP). In the RS state, the characteristic energy $\Omega$ and the system length $L$ are related via 
\begin{equation}
    -{\rm{ln}}{\Omega} \sim \sqrt{L},
    \label{eq:IRFP-scale}
\end{equation}
which indicates a formally infinite dynamical exponent. Moreover, the terminology \emph{infinite randomness} means that as the system is coarse grained and the characteristic energy scale decreases, the distributions of the logarithms of the magnitudes of the terms in the renormalized Hamiltonian become arbitrarily broad. As a results, the ratio of the magnitude of any two terms approaches either zero or infinity\cite[]{Motrunich.PRB.2000,Hida.JPSJ.1996}. 

However, similar RS state has been elusive in the higher dimensions for decades as SDRG breaks down\cite{Patric.prl.1982,lin.prb.2003}. In various 2D disordered systems, numerical renormalization group (RG) methods have indicated that the width of the distribution of the logarithms of the coupling strengths is finite.\cite[]{Patric.prl.1982,lin.prb.2003}. Furthermore, numerical simulations of spin-1/2 Heisenberg antiferromagnets on a square lattice with random exchange or random site dilution reveal the AF order vanishes only in the limit of infinite randomness or weakly remained, respectively\cite[]{Sandvik.prl.1995,Wessel.prb.2006,Sandvik.prb.2002}. 

On the other hand, the so-called quantum spin-liquid (QSL) state has been identified in various candidate materials like organic charge-transfer salt, synthetic herbertsmithite and recently discovered $\rm{YbMgGaO_4}$\cite[]{Li.SciRep.2015,Watanabe.JPSJ.2014,Freedman.JACS.2010}. While the power-law behavior observed in heat capacity and magnetic susceptibility in these materials suggests the presence of a QSL state with a spinon-Fermi surface \cite[]{Chen.prb.2016,Chen.prb.2017}, similar results have also been obtained in numerical simulations of related lattice models after introducing randomness \cite[]{Kawamura.JPSJ.2014,Kawamura.JPSJ.2017}. In Ref.~\cite[]{kimchi.prx.2018}, Kimchi et al. proposed that a network of spinons, emerging as topological defects between domain walls, could arise from a background of valence-bond solid (VBS) upon the introduction of randomness to the clean system. Then, the random-singlet like behavior could appear over a parametrically large range of length scales if the initial coupling distribution is parametrically broad and finally a spin-glass state would freeze at the lowest energies. 

Recently, Liu et al. \cite[]{Liu.Sandvik.2018.PRX} demonstrated the absence of antiferromagnetic (AFM) long-range order in the random $Q$ model on a square lattice by extrapolating the order parameter to the thermodynamic limit (TDL) using large-scale Monte Carlo simulations. 
Additionally, they found that the mean spin correlation decays as $1/r^2$, which corresponds to the properties of the IRFP obtained through the SDRG elimination process in the 1D RS. Similar behavior in mean spin correlation is also observed in RS-like phases within frustrated spin systems \cite[]{Ren.arXiv.2020}. However, unlike the infinite dynamical exponent $z$ in the 1D RS, a finite $z$ dependent on the strength of the randomness is determined through scaling analysis of heat capacity and uniform susceptibility versus temperature or system size\cite[]{Liu.prb.2020,Liu.Sandvik.2018.PRX}, which is similar to the Griffiths fixed point\cite[]{Griffiths.PRL.1969}.

In this work, to address the aforementioned controversy, we investigate the RS-like state in the random $Q$ model by directly extracting the singlet-triplet gap $\Delta_s$ using stochastic series expansion (SSE) method by employing restricted subspace update algorithm. Our numerical results reveal the following: (a) The scaling relation between energy and length scales is comparable to Eq.\ref{eq:IRFP-scale} within the system sizes we simulate. (b) Those scaling relations across different strengh of randomness $\Lambda$ can be collapsed onto a single curve by rescaling the system size $L$ with $L\Lambda^\alpha$. (c) The distribution width of the logarithms of the gap ${\rm ln}\Delta_s$, characterized by the standard deviation $\sigma$, diverge as the system size increases. Therefore, our numerical results suggest a single IRFP in the 2D RS phase of the random $Q$ model.

\section{Model Hamiltonian and method} 

\begin{figure}[t]
\centering
\includegraphics[width=0.4\textwidth]{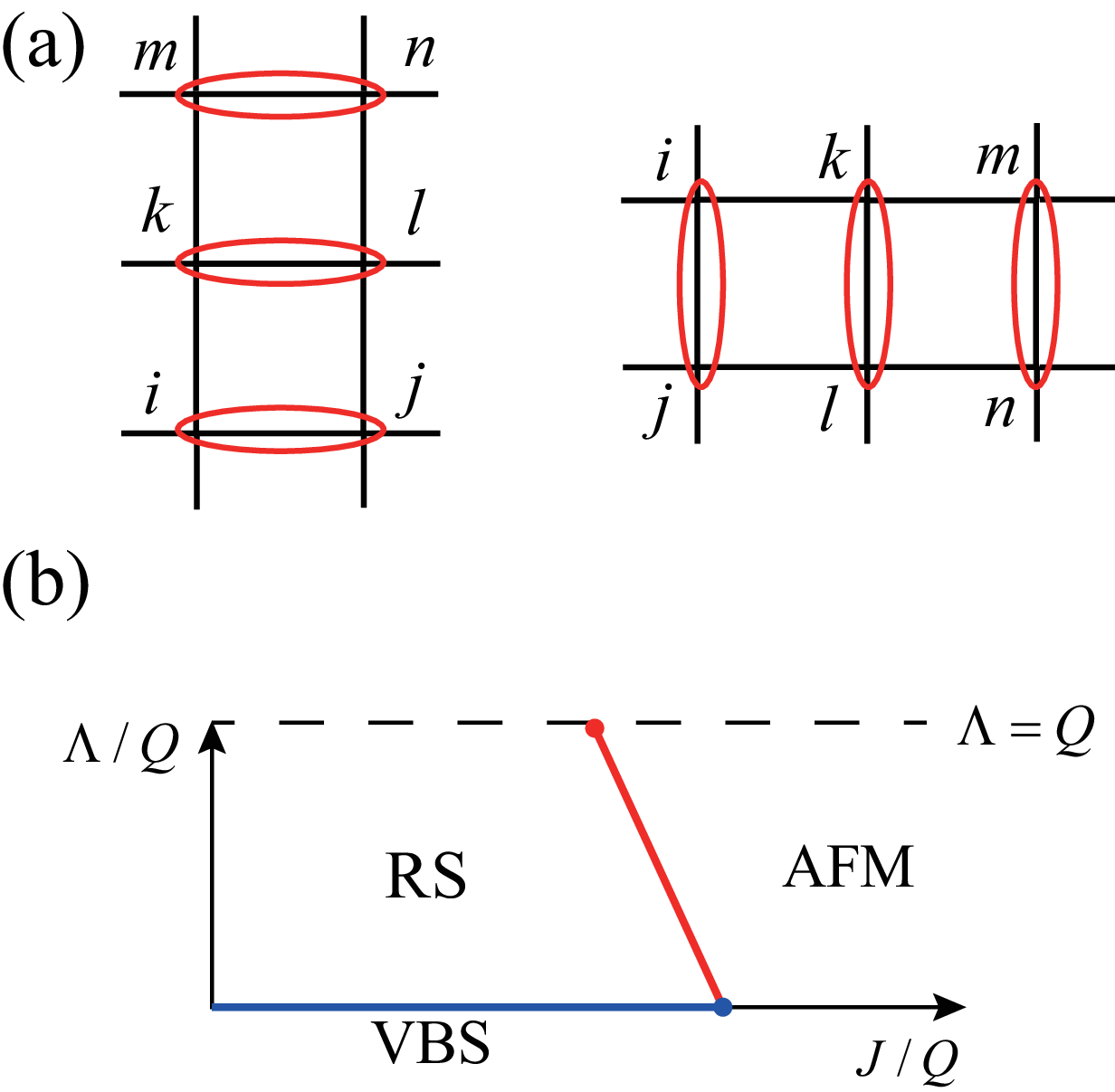}
\caption{(a) Patterns of the multispin interactionsred $Q$ term. (b) Phase diagram of the model.~\ref{eq:mdl}. The solid 
$x$-axis indicate the VBS-AFM transition in the pure $JQ_3$ model with $\Lambda=0$. The dashed line indicate the RS-AFM transition with $\Lambda=Q$, as studied in Ref.~\cite[]{Liu.Sandvik.2018.PRX}. The red line marks a continuous quantum phase transition between the AFM and RS states for at least some range of $\Lambda>0$. Since VBS is destroyed for any $\Lambda$, the RS state exists throughout the parameter space to the left of the phase boundary.}
\label{fig:lat}
\end{figure}

The model study in this work is based on the random $Q$ model on the squared lattice:
\begin{equation}
    {H} =  - J\sum\limits_{\left\langle {ij} \right\rangle } {{P_{ij}}} - \sum\limits_{\left\langle {ijklmn} \right\rangle } {{Q_{i}}{P_{ij}}{P_{kl}}{P_{mn}}},
    \label{eq:mdl}
\end{equation}
where $P_{ij}$ is singlet projectors on two nearest-neighbor $S=1/2$ spins $i,j$ and six-spin interaction $Q$ term is the product of three singlet projectors that forming a horizontal or vertical column as shown in Fig.~\ref{fig:lat}(a). Here, we choose $Q$ as energy unit for convenience, and the randomness is introduced by the following bimodel coupling distribution of the $Q$ interaction:
\begin{equation}
\pi(Q_i)=\begin{cases}
 1/2 & \text{ if } Q_i=Q+\Lambda, \\ 
 1/2 & \text{ if } Q_i=Q-\Lambda.
\end{cases}
\label{eq:bimodel-distribution}
\end{equation}
The phase diagram is denoted by the strength of randomness $\Lambda/Q$ and parameter $g=J/Q$ as shown in Fig.~\ref{fig:lat}(b). The $x$-axis denote the 1D phase diagram of pure system with the strength of randomness $\Lambda=0$, where parameter $g$ drives the AFM-VBS deconfined quantum-critical (DQC) point\cite{Lou.Sandvic.2009.PRB,Sentil.Ashvin.2004.Science}. The upper dashed line with $\Lambda=Q$ is carefully studied in the Ref.~\cite{Liu.Sandvik.2018.PRX}. In that work, instead of weak AFM, Liu et al. established the RS-AFM transition at $J_c/Q\sim0.81$. So, together with the fact that the VBS phase (marked by the solid blue line) would break for any $\Lambda>0$, there should be a phase boundary marked by a red line, which connect the DQC point and RS-AFM transition point, that seperate the RS and the AFM phase.

In order to investigate the scaling behavior of the RS state across different strengths of randomness $\Lambda$, we focus on the parameter line with $J=0$ along the $y$ axis as shown in Fig.~\ref{fig:lat}(b). The model.~\ref{eq:mdl} perserves spin SU(2) symmetry, allowing us to take the Monte Carlo sampling within the subspace with the spin $S_z=\mathrm{odd}$ and determine the lowest energy $E_{gs}^{s_z=odd}$ in these subspace. By subtracting the ground state energy $E_{gs}$ obtained via the standard SSE mothed, we derive the excitation gap $\Delta_s$, which represents the energy required for a spin-flip process in the ground state. It's important to note that, as discussed in Ref.~\cite[]{kimchi.prx.2018}, there exist other energy scales in this random $Q$ model describing the valence-bond type excitations $\Delta_{VB}$. Thus, the energy $\Delta_s$ calculated using our constrained subspace algorithm may not represent the true excitation gap of the model. Nevertheless, it specifically characterizes the excitation gap of the RS state we tend to study here, which is associated with the energy scale of singlet valence bond formation between pairs of local moments within the defect of the VBS background. Our constrained subspace algorithm has been successfully applied in studing the plaquette Heisenberg model in Ref.~\cite{Xiong.PRB.2022}.

In the following section, we calculate the singlet-triplet gap in the logarithm scale $\langle{\rm{ln}}\Delta_s\rangle$ and its standard deviation $\sigma=\sqrt{\langle ({\rm{ln}}\Delta_s-\langle {\rm{ln}}\Delta_s\rangle)^2\rangle}$ of the random $Q$ model described by Eq.~\ref{eq:mdl}. The braket $\langle...\rangle$ stands for the average among samples with different disorder realizations described by Eq.~\ref{eq:bimodel-distribution}. We set $Q=1$ as the energy unit and simulate system with different strengh of randomness $\Lambda$. The system size is choosen as $L_x=L_y=L$, where $L_x$ and $L_y$ is the unit cell along $x$ and $y$ direction, respectively. In this finite temperature simulation, we set $\beta=10L$ as the effective dynamical exponent $z$ is large andthe number of prepared samples increased from 256 to 640 as the system size $L$ grew from 8 to 32, in order to ensure the accuracy of the results. 

The singlet-triplet gap can be also obtained using the projector Monte-Carlo algorithm\cite{Sandvik.PRL.2005,Sandvik.PRB.2010}. However, its sampling efficiency becomes a significant limitation, particularly in scenarios involving a large amplitude of the $Q$-term, which is precisely the case we aim to investigate.  

\section{Numerical results} 

\subsection{\label{sec-a}Scaling behavior of the singlet-triplet gap with system size}
\begin{figure}[t]
    \centering
    \includegraphics[width=0.54\textwidth]{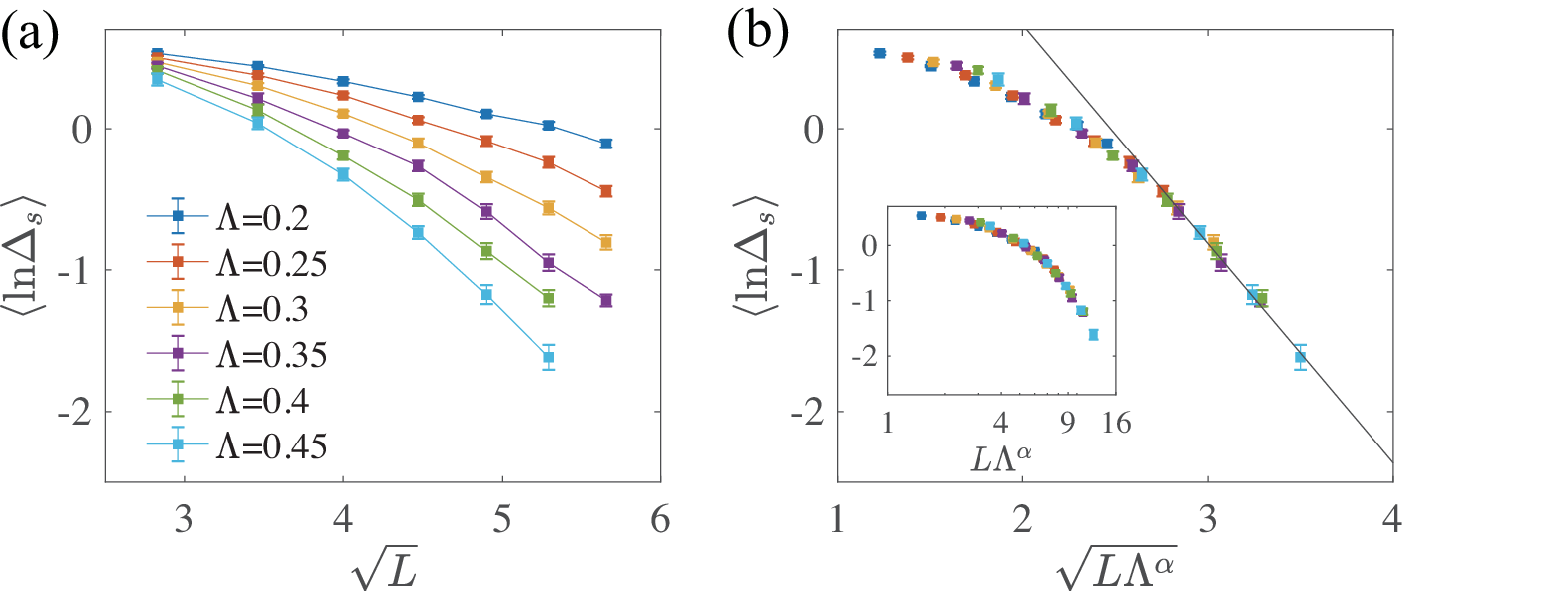}
    \caption{(a) The singlet-triplet gap $\Delta_s$ for systems with different strengths of randomness $\Lambda=0.2,0.25,0.3,0.35,0.4$ and $0.45$ versus different system size $L=8,12,16,20,24,28$ and $32$. We attempt to collapse the data of singlet-triplet gap $\Delta_s(\Lambda,L)$ into a single curve by rescaling the system size to $L^{\ast}=L\Lambda^{\alpha}$. The results are shown in sqrt scales (b) and log-log scales (inset). And the scaling exponent $\alpha$ is estimated to be $1.03(2)$.}
    \label{fig:gap-L}
\end{figure}

As we have mentioned above, compared with 1D RS, the similar properties observed in the random $Q$ model, such as the power law behavior of uniform susceptibility and heat capacity, as well as the $1/r^2$ behavior of the mean spin correlation function, were considered evidence that we found the RS phase in 2D\cite[]{Liu.prb.2020,Liu.Sandvik.2018.PRX,Kawamura.JPSJ.2014,Kawamura.JPSJ.2014}. And then, the dynamical exponent $z$ was estimated by fitting the scaling behavior of the uniform susceptibility $\chi\propto T^{D/z-1}$. However, in the previous study of model.~\ref{eq:mdl} with parameter $J=1$ and $\Lambda=Q$, it was found that the value of $z$ is finite and depends on the model parameter $Q/J$. These results are not easily understood, as both the inifinite $z$ and the quadratic behavior of mean spin correlation function are the typical characteristics of the IRFP obtained by the SDRG in 1D\cite[]{Liu.Sandvik.2018.PRX}. 

In this work, we aim to study the scaling relation of energy and length scales by directly calculating the singlet-triplet gap $\Delta_s$. This approach is necessary and important for several reasons: (a) it allows us to avoid logarithmic corrections, such as the uniform susceptibility, $\chi(T)\propto \frac{1}{T}{\rm{ln}}T$ in the 1D RS state, which can significantly affect the extracted value of $z$ when fitting $\chi\propto T^{D/z-1}$, and (b) instead of the valence bond type excitation $\Delta_{VB}$ in the VBS background, the extracted singlet-triplet gap represents the true energy scale of the spinon network that we are intrested in. 

Here, we calculate the singlet-triplet gap $\Delta_s$ for different strengths of randomness $\Lambda$ and system sizes ranging from $L=8$ to $32$. To compare with the scaling relation Eq.~\ref{eq:IRFP-scale}, we first present the data in log-sqrt plot, as shown in Fig.~\ref{fig:gap-L}(a). We exclude data with both large strength of randomness and large system size because it is not reliable in our calculations due to the divergence of the width of the gap distributions, which will be discussed in the next subsection. In previous studies, it was considered that the dynamical exponent $z$ in the 2D RS is finite and dependent on the randomness strength in the Hamiltonian \ref{eq:mdl}. It was suggested that these 2D RS states, which depend on the model parameters, correspond to different fixed points, similar to the Griffiths fixed point \cite{Liu.Sandvik.2018.PRX}. However, in our direct calculation of the gap and plotting it on a log-sqrt scales, the data of $\Delta_s(\Lambda,L)$ with different system size $L$ seems share similar monotonically asymptotic behavior in the tail of each curve with different strengths of randomness $\Lambda$. Therefore, we attempt to collapse the data with different $\Lambda$ onto a single curve, $\Delta_s(\Lambda,L)=f(L^{\ast})$, by rescaling the system size according to 
\begin{equation}
    L^{\ast} = L\Lambda^{\alpha},
    \label{eq:rescale}
\end{equation}
which is similar to the approach used in the study of the diluted Heisenberg model\cite[]{Wang.PRB.2010}. This assumption suggests that the strength of randomness $\Lambda$ may effectively play a role analogous to the system size $L$ in the random $Q$ model. Considering that the true length scale of the RS state in the random $Q$ model should correspond to the size of the spinon subsystem, which arises as a defect within the VBS domain wall. Since randomness can create VBS domains within the intact VBS background, depending on its strength, it is reasonable to use a general scaling function Eq.~\ref{eq:rescale} to define an effective length scale for the spinon subsystem. Our assumption is supported by the nice datacollapse achieved with an optimized scaling exponent $\alpha=1.03(2)$ as shown in Fig.~\ref{fig:gap-L}(b). 

In the log-sqrt scale, the scaling relation described by Eq.~\ref{eq:IRFP-scale} does not appear to hold at smaller effective size $L^{\ast}$. This result is reasonable as RS state behavior emerges only at lower energy scales after decimating a significant number of strong bonds in the RG process. Compared to the ${\rm{ln}}\Delta_s$ versus $L$ for different randomness strengths $\Lambda$ shown in Fig.~\ref{fig:gap-L}(a), a clear linear relation of ${\rm{ln}}\Delta_s$ with the rescaled system size $\sqrt{L^{\ast}}$ is observed for large $\sqrt{L^{\ast}}$, indicating that RS physics manifests at this length scale. Our results suggest that, at least within the range of length scales we simulate, the energy and length scales appear to follow the 1D RS relation described by Eq.~\ref{eq:IRFP-scale}. Although scaling behavior can vary during the RG flow, the similar IRFP behavior observed in our simulations help explain why the $1/r^2$ decay of the mean spin correction function, a typical outcome of the IRFP, is present in the random $Q$ model.


Before proceeding, we note that when the data are plotted in the log-log scale [see the inset of Fig.~\ref{fig:gap-L}(b)], the slope corresponding to the dynamical exponent $z$ increases with the rescaled system size $L^\ast$. This is consistent with the increase of $z$ found in Ref.~\onlinecite{Liu.prb.2020} with the staggered susceptibility and the spin structure factor, in which $z$ was argued to approach a finite value in the thermodynamic limit. However, based on the log-sqrt plot and the fitting shown above, we believe that the increasing slope in the log-log plot is a signature of an infinite dynamical exponent $z$ instead.

\subsection{\label{subsect-B}Scaling behavior of the deviation of the logarithm of the singlet-triplet gap with system size}
\begin{figure}[t]
    \centering
    \includegraphics[width=0.54\textwidth]{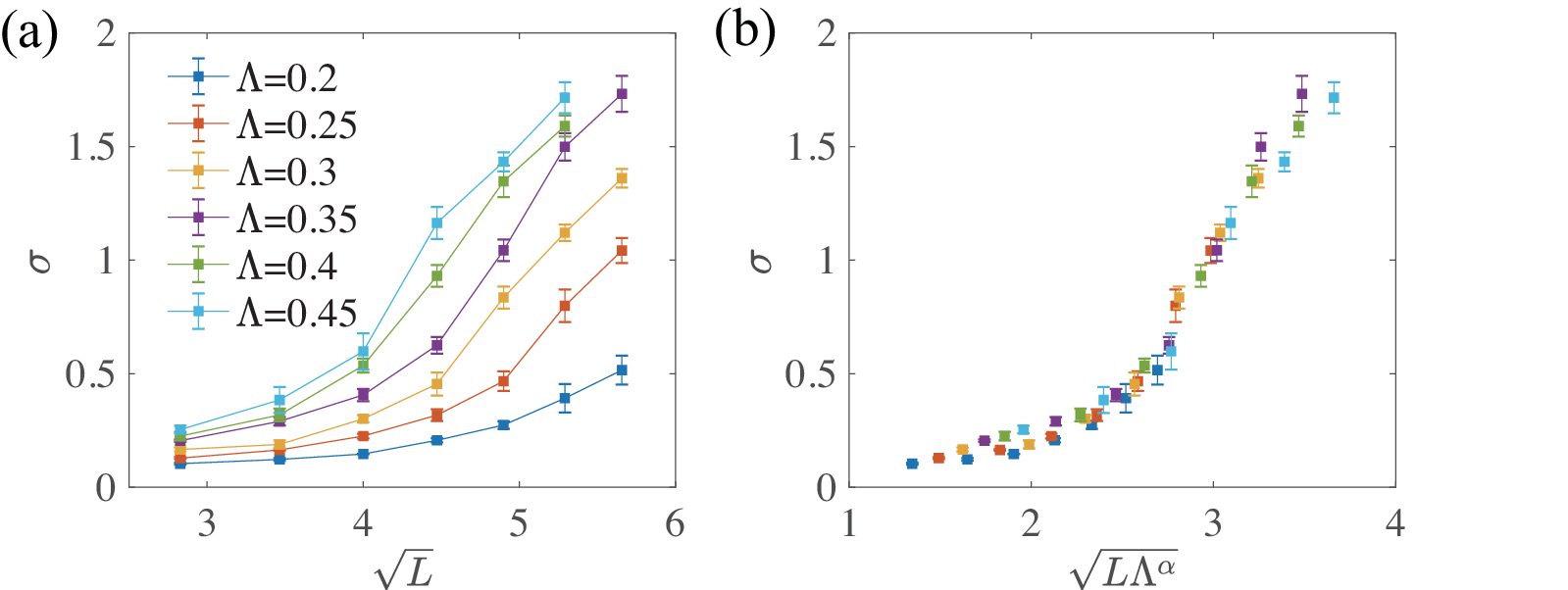}
    \caption{(a) The deviation $\sigma(\Lambda,L)$ of the logarithm of the excitation gap ${\rm{ln}}\Delta_s$ for different model parameters $\Lambda$ and system size $L$. (b) The datacollapse of $\sigma(\Lambda,L)$ by rescaling the system size to $L^{\ast}=L\Lambda^{\alpha}$. The scaling exponent $\alpha$ is estimated to be $0.92(4)$.}
    \label{fig:dev}
\end{figure}

Another advantage of directly calculating the singlet-triplet gap is that it allows us to examine the scaling behavior of the deviation of ${\rm{ln}}\Delta_s$ with respect to system size $L$, which is crucial for studing the RS state in the random $Q$ model. As it well known, previous difficulties in identifying the 2D RS state are attributed to the breakdown of the SDRG method. In other words, the absence of divergence in the deviation of the logarithmic distribution of single-triplet gap with system size, as found in earlier numerical RG studies, indicates that there is no IRFP in 2D quantum magnets with SU(2) spin-isotropic interactions\cite{Patric.prl.1982, lin.prb.2003}. Therefore, it is essential to analyze the scaling behavior of $\sigma=\sqrt{\langle ({\rm{ln}}\Delta_s-\langle {\rm{ln}}\Delta_s\rangle)^2\rangle}$ and to determine whether the RS behavior in the random $Q$ model follows the relation $\sigma\propto\sqrt{L}$, which is considered a fingerprint of the IRFP\cite[]{Hida.JPSJ.1996,Fisher.PRB.1994}.

To analyze how the width of the distribution of ${\rm{ln}}\Delta_s$ changes and to compare it with the 1D RS behavior, we examine the deviation $\sigma$ as a function of the system size $\sqrt{L}$. Our results indicate that $\sigma$ remains relatively stable for small system sizes and weak disorder but increases significantly as the system size and disorder strength grow, as shown in Fig.~\ref{fig:dev}(a). The scaling relation $\sigma\propto\sqrt{L}$ appears to hold in the region with larger $\Lambda$ and larger system sizes $L$, supporting our previous finding that the IRFP can describe the physics within the range of length scales we simulate. It is essential to note that that the width of the distribution of ${\rm{ln}}\Delta_s$ broadens with increasing $\Lambda$ and $L$, suggesting that the single-triplet gap may become very small in some samples under these parameter. Due to the very small values of $\Delta_s$, our simulation precision cannot provide reliable $\sigma$ for $\Lambda = 0.45$ and $0.5$ with system size $L = 32$. Consequently, these $\sigma(\Lambda,L)$ values deviate significantly from their expected asymptotic behavior with $L$ and are not shown in the Fig.~\ref{fig:dev}. However, this does not affect the overall qualitative conclusions of our study.

Finally, we apply the same rescaling form from Eq.~\ref{eq:rescale} to collapse the $\sigma(\Lambda,L)$ data onto a single curve. The result, shown in Fig.~\ref{fig:dev}(b), is obtained with an optimized scaling exponent of $\alpha=0.92(4)$, consistent with the value $\alpha=1.03(2)$ obtained from the datacollapse of $\Delta_s$. As noted in the previous section, RS physics described by the IRFP also manifests at large effective sizes $L^{\ast}$. The distinct kink observed in the datacollapse of $\sigma(\Lambda,L)$, similar to that seen in the datacollapse of the $\Delta_s$ shown in Fig.~\ref{fig:gap-L}(b), may indicate different stages of the RG process. 

\section{Summary and discussion}
In this study, we develop the constrained subspace update algorithm within the SSE Monte-Carlo method to calculate the singlet-triplet excitation gap $\Delta_s$ of the random $Q$ model. By directly computing the singlet-triplet excitation gap, we can not only observe the scaling relations of energy and length scales directly, but also avoid the influence of valence-bond type excitations on these scaling relations.

We numerically simulate the singlet-triplet excitation gap and the deviation $\sigma$ of the logarithm of this gap distribution in the random $Q$ model, in which its ground state is identified as the 2D RS. Our numerical simulations reveal that the 2D RS state in the random $Q$ model exhibits a similar energy-length scaling relation, up to a system size of $L=32$, to the one observed in the 1D RS state described by the IRFP derived using SDRG. This finding helps reconcile the contradiction observed in previous studies: the decay of the mean spin correlation as $1/r^2$ and the presence of a finite dynamical exponent $z$ in the 2D RS state. And the reason why the scaling relations ${\rm{ln}\rm{\Delta_s}}\propto \sqrt{L}$ and $\sigma \propto \sqrt{L}$ hold (at least for a range of length scales) in 2D quantum magnets with SU(2) spin-isotropic interactions remains an open question, which is beyond the scope of this work.

The data collapse with the rescaled system size $L^{\ast}=L\Lambda^{\alpha}$ with an exponent $\alpha\simeq 1$ deserves further discussions. While the low-energy spin degrees of freedom are embedded in the VBS domain walls and vortices as shown in previous works \cite{kimchi.prx.2018, Liu.Sandvik.2018.PRX, Liu.prb.2020}, the characteristic length scale in the RS phase is given by the typical size $\ell$ of VBS domains. Given that the spatial quenched disorder acts as a random field linearly coupled with the VBS order parameter, the typical domain size may be estimated following the Imry-Ma argument \cite{Imry1975}. If the two-component VBS order parameter has an enlarged O(2) continuous symmetry in the RS phase, the domain wall energy cost due to continuously rotating VBS order parameters scales as $O(\ell^{d-2})\sim O(1)$ ($d=2$ is the spatial dimension). On the other hand, the effective random field summed in one VBS domain fluctuates as $\langle \Delta h^{2}\rangle\sim O(\Lambda^{2}\ell^{d})$, thus the energy gain in forming domains scales as $O(\Lambda\ell)$. The energy balance in forming VBS domains leads to the typical domain size $\ell\sim \Lambda^{-1}$, and thus the rescaled system size is given by $L^{\ast}=L/\ell\sim L\Lambda$. This implies the exponent $\alpha=1$, which is consistent with our numerical results.

Moreover, considering that the spinon network emerges as a subsystem within the nexus of the domain wall of the VBS background, we propose that the strength of randomness $\Lambda$ could scale the effective system size as $L\Lambda^{\alpha}$. The datacollapse of both $\Delta_s(\Lambda,L)$ and $\sigma(\Lambda,L)$ supports our hypothesis. These results suggest that the RS state appearing in the phase diagram Fig.~\ref{fig:lat}(b) might belong to the same fixed point.

\begin{acknowledgments}
    C.P. is grateful to helpful discussions with R.-Z.Huang and
    H.Shao. This work is supported by the National Natural Science Foundation of China (Grant Nos. 12174387 and 12304182), Chinese Academy of Sciences (Nos. YSBR-057, JZHKYPT-2021-08 and XDB28000000), and the Innovative Program for Quantum Science and Technology (No. 2021ZD0302600). 
\end{acknowledgments}

\bibliography{ref}
\end{document}